\documentclass[aps,prl,twocolumn,showpacs,preprintnumbers,amsmath,amssymb,superscriptaddress]{revtex4}%

\usepackage{graphicx}
\usepackage{dcolumn}
\usepackage{bm}
\usepackage{color}

\begin{document}

\preprint{preprint(\today)}

\title{Probing the quantum phase transition in Mott insulator BaCoS$_{2}$ tuned by pressure and Ni-substitution}

\author{Z.~Guguchia}
\affiliation{Department of Physics, Columbia University, New York, NY 10027, USA}
\affiliation{Laboratory for Muon Spin Spectroscopy, Paul Scherrer Institute, CH-5232
Villigen PSI, Switzerland}

\author{B.A. Frandsen}
\affiliation{Department of Physics and Astronomy, Brigham Young University, Provo, Utah 84602, USA}

\author{D. Santos-Cottin}
\affiliation{IMPMC, Sorbonne Universit\`{e}s-UPMC, CNRS, IRD, MNHN, 4, place Jussieu, 75005 Paris, France}

\author{S.C. Cheung}
\affiliation{Department of Physics, Columbia University, New York, NY 10027, USA}

\author{Z. Gong}
\affiliation{Department of Physics, Columbia University, New York, NY 10027, USA}

\author{Q. Sheng}
\affiliation{Department of Physics, Columbia University, New York, NY 10027, USA}

\author{K. Yamakawa}
\affiliation{Department of Physics, Columbia University, New York, NY 10027, USA}

\author{A.M. Hallas}
\affiliation{Department of Physics and Astronomy, McMaster University, Hamilton, ON L8S 4M1 Canada}

\author{M.N. Wilson}
\affiliation{Department of Physics and Astronomy, McMaster University, Hamilton, ON L8S 4M1 Canada}

\author{Y. Cai}
\affiliation{Department of Physics and Astronomy, McMaster University, Hamilton, ON L8S 4M1 Canada}

\author{J. Beare}
\affiliation{Department of Physics and Astronomy, McMaster University, Hamilton, ON L8S 4M1 Canada}

\author{R.~Khasanov}
\affiliation{Laboratory for Muon Spin Spectroscopy, Paul Scherrer Institute, CH-5232
Villigen PSI, Switzerland}

\author{R.~De Renzi}
\affiliation{Department of Mathematical, Physical and Computer Sciences, Parco delle Scienze 7A, I-43124 Parma, Italy}

\author{G.M. Luke}
\affiliation{Department of Physics and Astronomy, McMaster University, Hamilton, ON L8S 4M1 Canada}
\affiliation{Canadian Institute for Advanced Research, Toronto, ON Canada M5G 1Z7}
\affiliation{TRIUMF, Vancouver, BC, Canada V6T 2A3}

\author{S. Shamoto}
\affiliation{Advanced Science Research Center, Japan Atomic Energy Agency (JAEA), Tokai, Naka, Ibaraki 319-1195, Japan}

\author{A. Gauzzi}
\affiliation{IMPMC UMR7590, Sorbonne Universit\`{e}, CNRS, IRD, MNHN, 4, place Jussieu, 75005 Paris, France}

\author{Y. Klein}
\affiliation{IMPMC UMR7590, Sorbonne Universit\`{e}, CNRS, IRD, MNHN, 4, place Jussieu, 75005 Paris, France}

\author{Y.J.~Uemura}
\email{yu2@columbia.edu}
\affiliation{Department of Physics, Columbia University, New York, NY 10027, USA}

\begin{abstract}

We present a muon spin relaxation study of the Mott transition in BaCoS$_{2}$ using two independent control parameters: (i) pressure $p$ to tune the electronic bandwidth and (ii) Ni-substitution $x$ on the Co site to tune the band filling. For both tuning parameters, the antiferromagnetic insulating state first transitions to an antiferromagnetic metal and finally to a paramagnetic metal without undergoing any structural phase transition. BaCoS$_2$ under pressure displays minimal change in the ordered magnetic moment $S_{ord}$ until it collapses abruptly upon entering the antiferromagnetic metallic state at $p_{cr}\sim$1.3 GPa. In contrast, $S_{ord}$ in the Ni-doped system Ba(Co$_{1-x}$Ni$_{x}$)S$_{2}$ steadily decreases with increasing $x$ until the antiferromagnetic metallic region is reached at $x_{cr} \sim 0.22$. In both cases, significant phase separation between magnetic and nonmagnetic regions develops when approaching $p_{cr}$ or $x_{cr}$, and the antiferromagnetic metallic state is characterized by weak, random, static magnetism in a small volume fraction. No dynamical critical behavior is observed near the transition for either tuning parameter. These results demonstrate that the quantum evolution of both the bandwidth- and filling-controlled metal-insulator transition at zero temperature proceeds as a first-order transition. This behavior is common to magnetic Mott transitions in $RE$NiO$_{3}$ and V$_{2}$O$_{3}$, which are accompanied by structural transitions without the formation of an antiferromagnetic metal phase.

\end{abstract}
\maketitle

\section{Introduction}

The Mott metal-insulator transition (MIT), known to provide a platform for emergent phenomena such as high-temperature superconductivity and colossal magnetoresistance, remains one of the most intensely studied topics in condensed matter physics  \cite{Imada1998,Keimer2015,Vojta,Torrance,McWhan,Uemura1984}. This transition can occur as a thermal phase transition or as a quantum phase transition (QPT) near zero temperature. In the latter case, the transition is typically controlled by varying the electronic bandwidth using hydrostatic or chemical pressure or by varying the band filling via chemical substitution. The Mott transition usually occurs from an antiferromagnetic insulator (AFI) phase to a paramagnetic metal (PMM)
phase and is often accompanied by a structural phase transition. In some Mott systems, a direct one-step transition from AFI to PMM phase is observed,  while  in others the transition occurs in two steps involving an intermediate antiferromagnetic metal phase (AFM).  In order to understand fully the physics of the Mott transition, the following challenges need to be met: (i) Disentangling the different contributions of the charge, magnetic, and structural  interactions; (ii) establishing whether the transition is first-order or continuous; (iii) clarifying the similarities and differences between the one-step and two-step transitions; and (iv) comparing the effects of bandwidth control versus filling control.

Here, we attempt to address these challenges by elucidating the mechanism of the quantum MIT in the quasi two-dimensional system Ba(Co$_{1-x}$Ni$_{x}$)S$_{2}$ \cite{Martinson, Mentink, Kodama, Shamoto, Yasui}, where the MIT can be controlled either by pressure or by the partial substitution of Co for Ni \cite{Kodama}. The crystal structure, shown in  Fig. 1(a), consists of alternately stacked side-sharing (Co,Ni)S$_{5}$  pyramids, where the Co and Ni ions form a square lattice. Previous studies have shown that the unsubstituted ($x$ = 0) BaCoS$_{2}$ compound is an antiferromagnetic insulator with a N\'{e}el temperature $T_{N} = 300$~K at ambient pressure. Neutron scattering studies \cite{Kodama} indicate that Co$^{2+}$ is in the high spin state with an ordered moment of ${\sim}$ 3 ${\mu}_{B}$, which is progressively reduced upon Ni substitution. As shown in the phase diagrams in Figs. 1(b) and (c), the Mott transition from AFI to PMM occurs at a critical pressure of $p_{cr} \sim 1.3$~GPa \cite{Yasui1999,Sato1998} or Ni concentration of $x_{cr} \sim 0.22$ \cite{Martinson,SantosD}. Contrary to the commonly studied systems $RE$NiO$_{3}$ ($RE$ = Rare Earth) and V$_{2}$O$_{3}$, the Mott transition of BaCoS$_{2}$ does not involve any structural distortions and it occurs in two steps with the formation of an intermediate AFM phase. Because only electronic degrees of freedom come into play, BaCoS$_{2}$ is a model system to investigate the Mott MIT. A study of the evolution of the magnetic properties across the MIT would enable a comparison with the previously studied $RE$NiO$_{3}$ and V$_{2}$O$_{3}$ systems, possibly clarifying the role of structural transitions and of the intermediate AFM phase in the Mott transition.

\begin{figure*}[t!]
\centering
\includegraphics[width=1.0\linewidth]{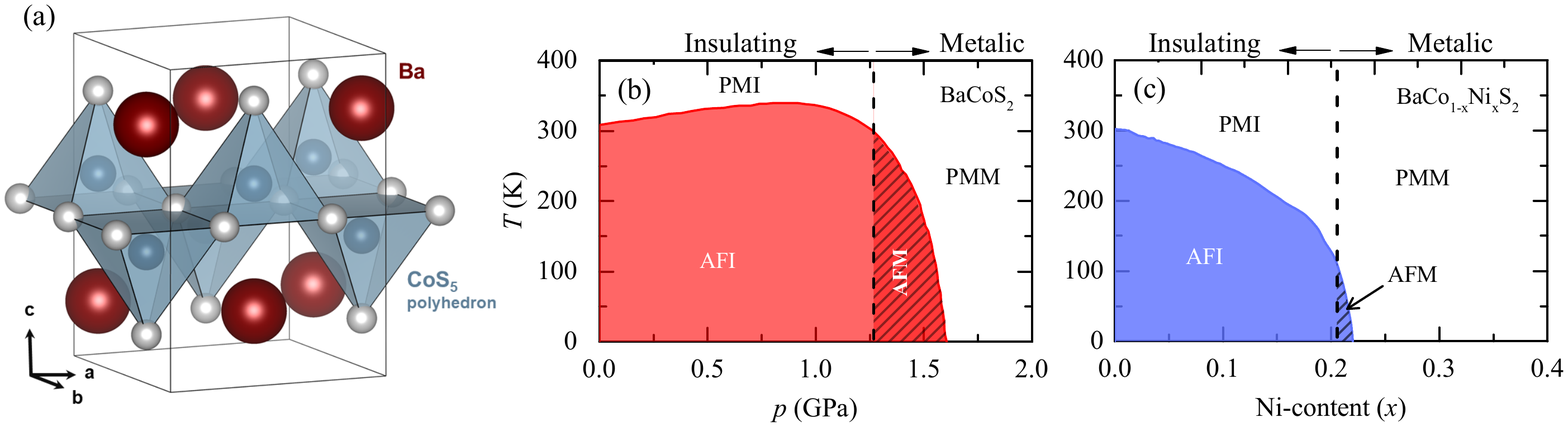}
\vspace{-6.8cm}
\caption{ (Color online) (a) Crystal structure of BaCoS$_{2}$. Schematic Temperature-Pressure (b) and Temperature-Doping (c) phase diagrams of BaCoS$_{2}$ and BaCo$_{1-x}$Ni$_{x}$S$_{2}$, respectively (adapted from Refs. [12,13]). PMI stands for paramagnetic insulator, AFI for antiferromagnetic insulator, PMM for paramagnetic metal, and AFM for antiferromagnetic metal. The conductive metal-insulator transition occurs at the broken vertical lines.} 
\label{fig1}
\end{figure*}

\begin{figure}[b!]
\centering
\includegraphics[width=2.8\linewidth]{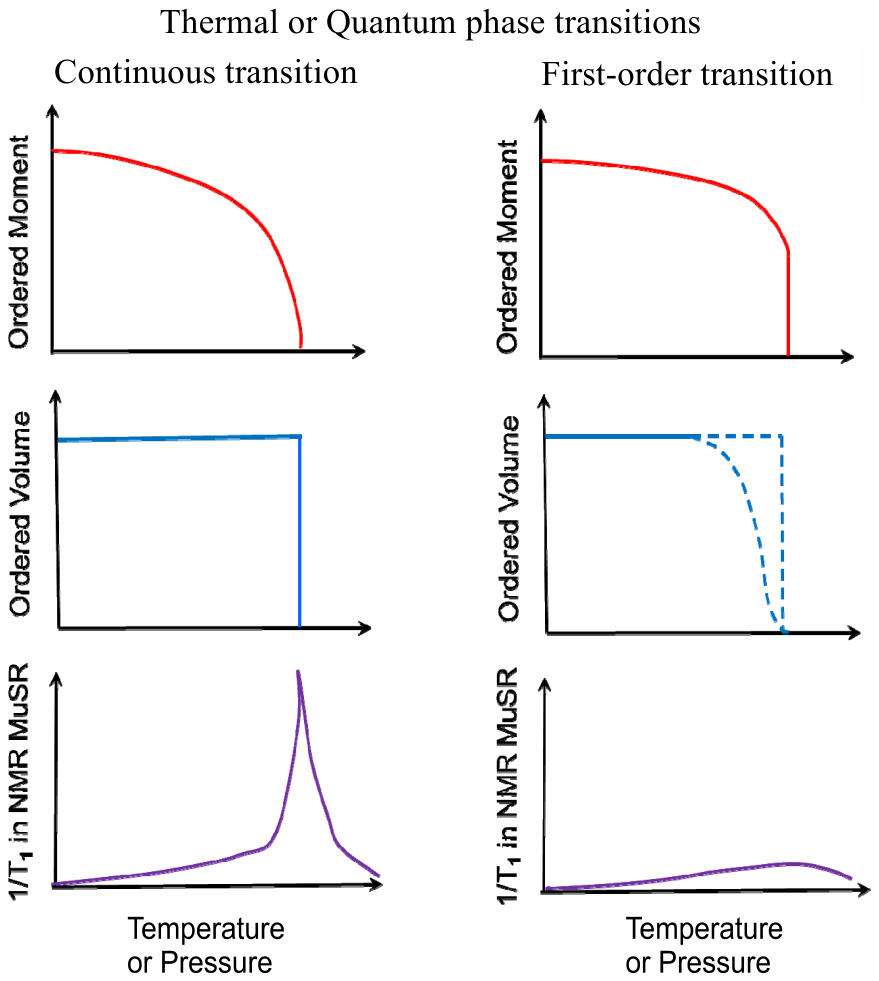}
\vspace{-7.0cm}
\caption{ (Color online) Schematic illustration of the thermal or quantum evolution of the ordered moment, magnetically ordered volume fraction, and relaxation rate in the case of continuous (left) and first-order (right) transitions.  Phase separation seen in the gradual change of the ordered volume fraction is a sufficient but not necessary condition for a first-order transition.} 
\label{fig2}
\end{figure}

\begin{figure}[t!]
\includegraphics[width=1.0\linewidth]{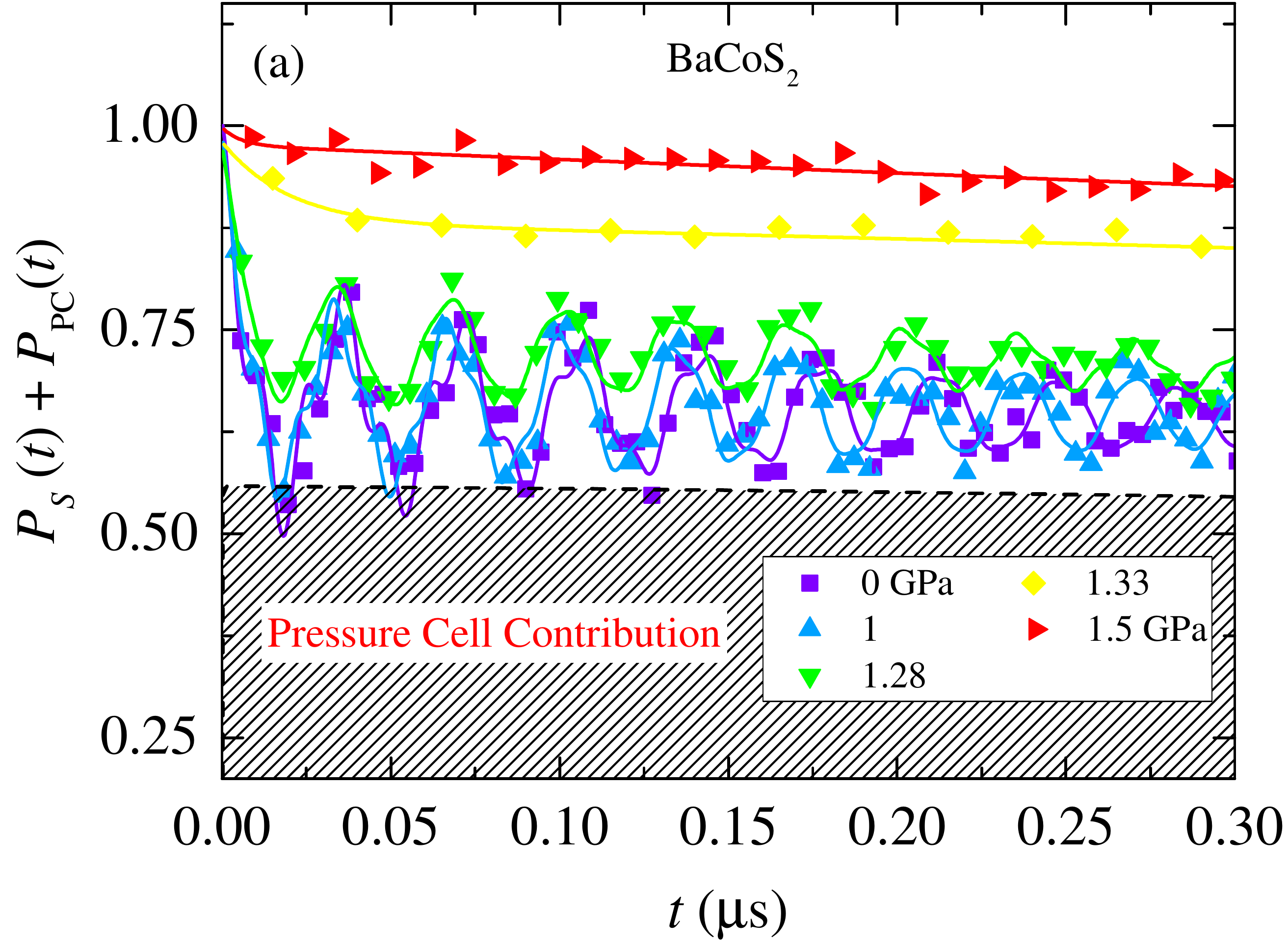}
\includegraphics[width=1.0\linewidth]{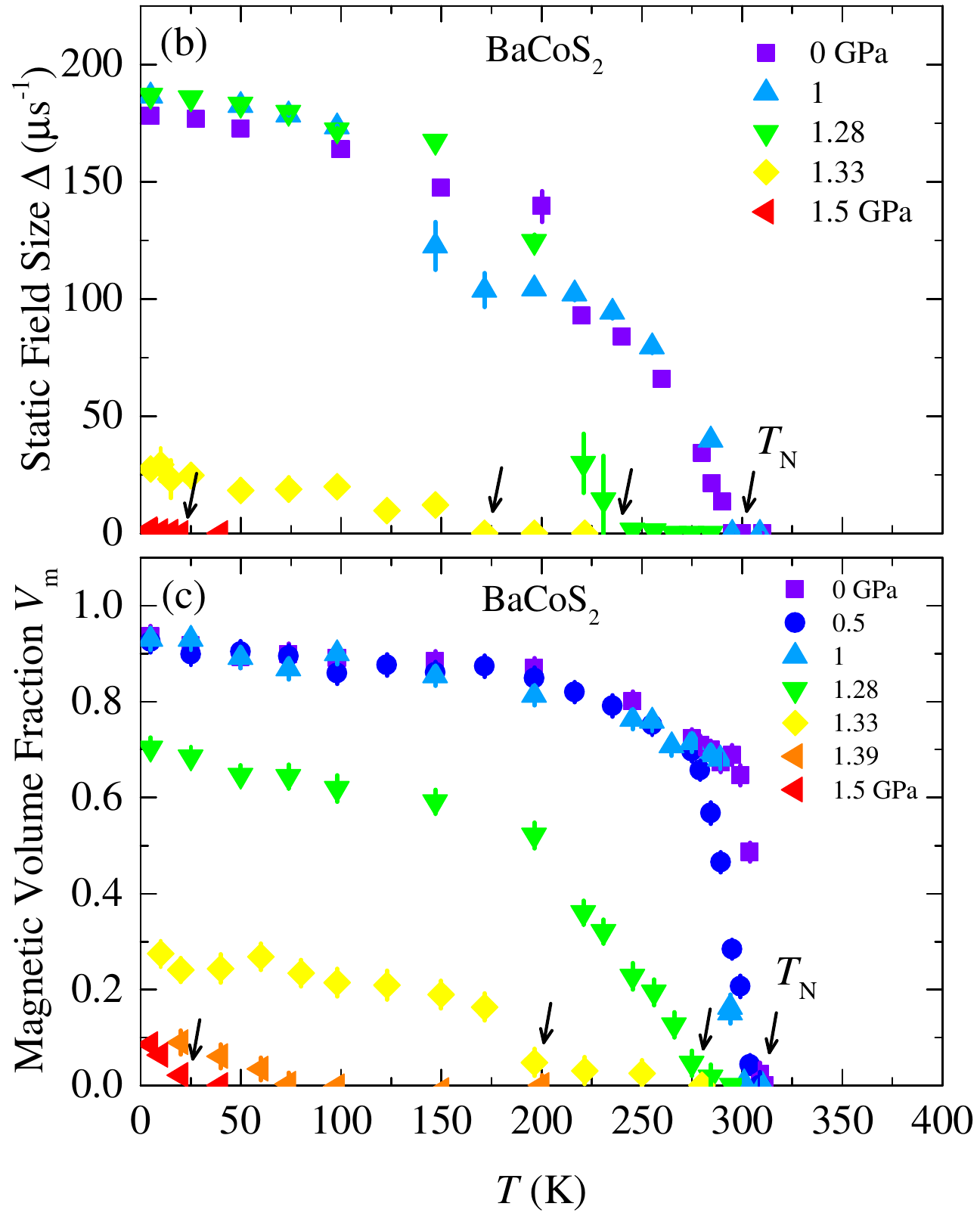}
\vspace{-0.5cm}
\caption{ (Color online) (a) Muon spin polarisation in zero field for BaCoS$_{2}$ recorded
at $T$ = 2 K under various applied pressures. The solid curves represent fits to the data by means of Eq. (2). (b) The temperature dependence of the static internal field size ${\Delta}(T)$=[(2${\pi}$${\nu}$)$^{2}$ + ${\lambda_{T}}$$^{2}$)$]^{1/2}$ under various applied hydrostatic pressures. The arrows mark the magnetic ordering temperature $T_{\rm N}$. (c) Same as (b), but displaying the magnetically ordered volume fraction.}  
\label{fig3}
\end{figure}

To achieve this, we have carried out a systematic muon spin relaxation (${\mu}$SR) study of the MIT in BaCoS$_{2}$ using $p$ and $x$ as control parameters for the bandwidth-tuned and filling-tuned transitions, respectively. ${\mu}$SR is an ideal probe for our purpose, since it can independently determine the magnitude of the local ordered moment and the volume fraction of the magnetically ordered regions, as well as detect dynamic magnetic critical behavior via measurements of the 1/$T_{\rm 1}$ relaxation rate. Since ${\mu}$SR is a point-like real-space magnetic probe which detects magnetic transitions in time space, we use the term "magnetic order" in this paper as implying static spin freezing while not referring to specific spatial spin correlation length. Taking advantage of these features, significant progress was made by recent ${\mu}$SR studies of the aforementioned prototypical Mott systems $RE$NiO$_{3}$ and V$_2$O$_{3}$ \cite{Frandsen2016}, where the substitution of $RE$ ions of different size and the application of hydrostatic pressure in V$_2$O$_3$ enable a bandwidth-controlled Mott transition, accompanied by a first-order structural transition. In both systems, the QPT from the AFI phase to the PMM phase occurs through a gradual reduction of the magnetically ordered volume fraction near the QPT until it reaches zero at the transition, while the magnitude of the ordered moment in the ordered regions of the sample remains unchanged until dropping abruptly to zero in the PMM state. No dynamical critical behavior occurs. These are typical features expected in the case of a first-order phase transition, as illustrated in Fig. 2. Such behavior was previously observed in weak itinerant magnets like MnSi and (Sr,Ca)RuO$_{3}$, where the magnetic transition is tuned by hydrostatic pressure \cite{UemuraMnSi} or the Ca/Sr substitution \cite{UemuraMnSi}. In MnSi, Fe substitution on the Mn site turns this transition to a continuous transition, as shown by a recent ${\mu}$SR experiment \cite{Goko} and a theoretical study \cite{Sang}.

In this paper, we show that the Mott QPT in BaCoS$_2$ is also first order for both pressure and Ni substitution. For both tuning parameters, the magnetically ordered volume fraction steady decreases near the AFI to AFM transition and reaches zero at the AFM to PMM transition, revealing a broad region of phase separation between magnetically ordered and disordered regions. In the case of pressure, the ordered moment retains its maximal value until an abrupt reduction occurs upon entering the AFM state, whereas the Ni substitution shows a more gradual decrease of the ordered moment with increasing Ni concentration. In both cases, the quantum evolution to the PMM state occurs without dynamical critical behavior. These observations lead to the conclusion that both the bandwidth- and filling-controlled Mott transitions in BaCoS$_2$ are first order. We also confirmed absence of static magnetism in the $x=1$ end-member compound BaNiS$_{2}$.

\begin{figure}[b!]
\centering
\includegraphics[width=1.0\linewidth]{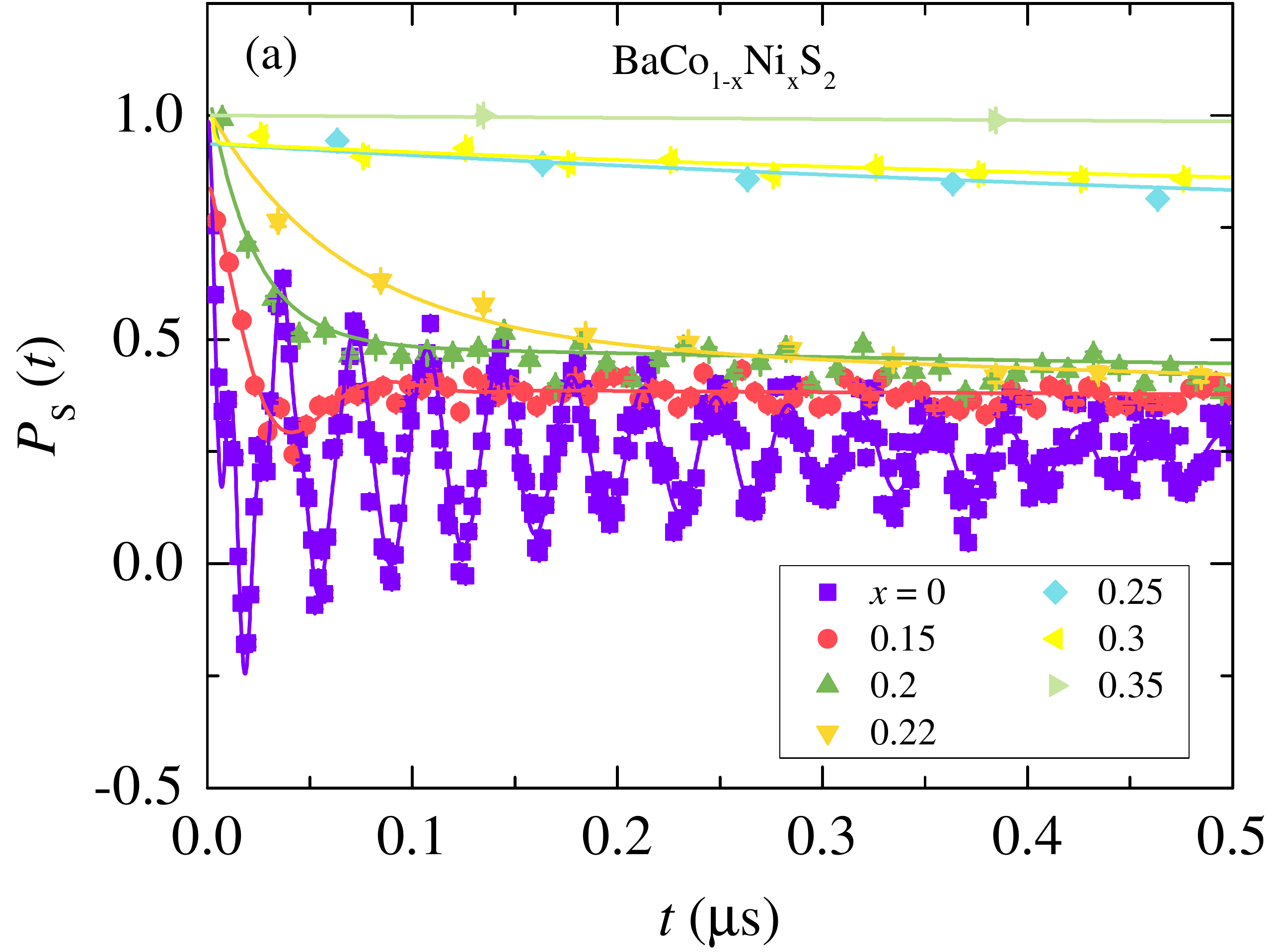}
\includegraphics[width=1.0\linewidth]{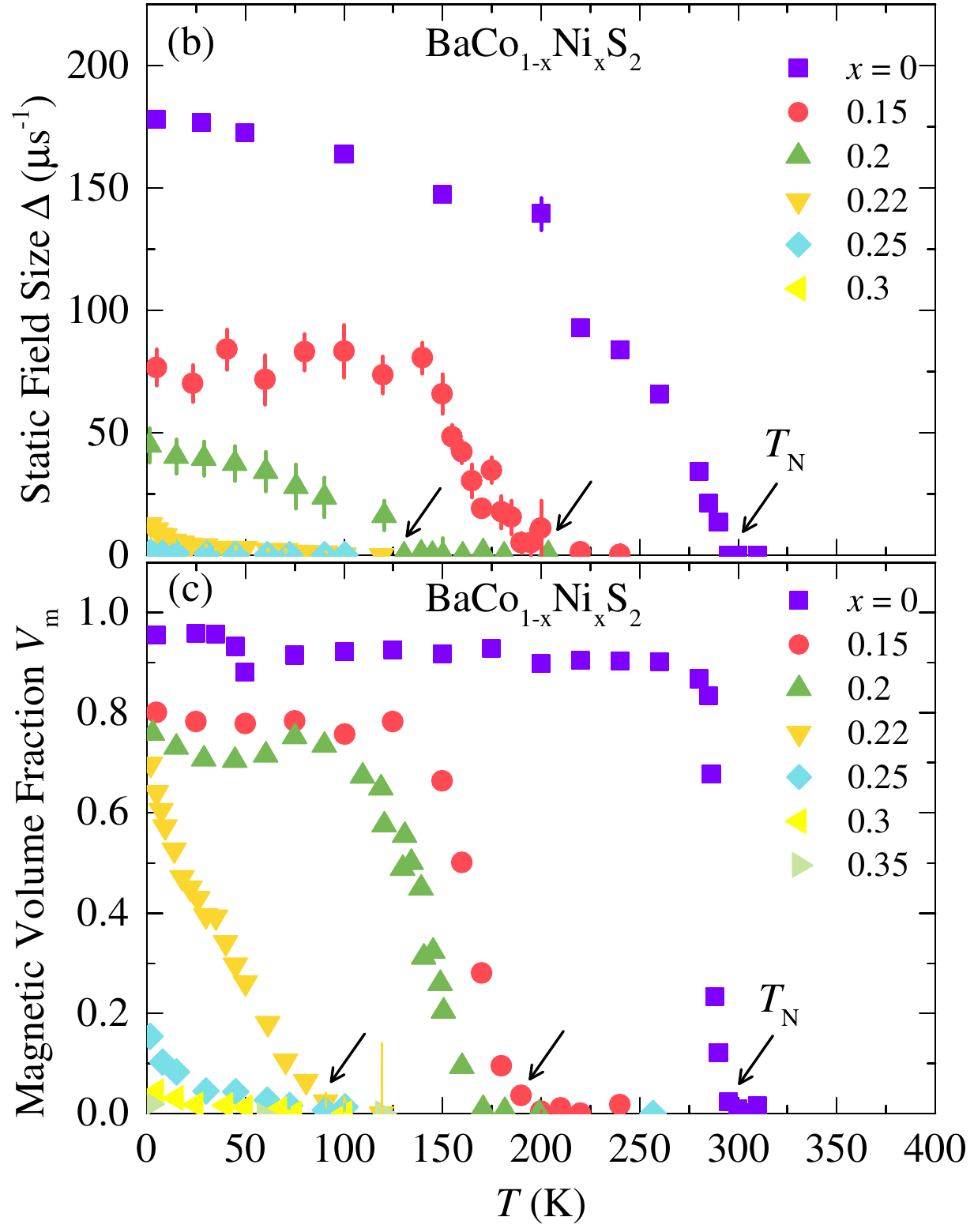}
\vspace{-0.7cm}
\caption{ (Color online) (a) Muon spin polarisation in zero field for Ba(Co$_{1-x}$Ni$_{x}$)S$_{2}$ recorded
at $T$ = 2 K for various values of $x$. The solid curves represent fits to the data by means of Eq. (2). (b) The temperature dependence of the static internal field size ${\Delta}(T)$=[(2${\pi}$${\nu}$)$^{2}$ + ${\lambda_{T}}$$^{2}$)$]^{1/2}$ for various values of $x$. The arrows mark the magnetic ordering temperature $T_{\rm N}$. (c) Same as (b), but displaying the magnetically ordered volume fraction.} 
\label{fig4}
\end{figure}

\section{Experimental Details}

Polycrystalline samples of Ba(Co$_{1-x}$Ni$_{x}$)S$_{2}$ were synthesized using a conventional solid state reaction method, as described in detail elsewhere \cite{Shamotogrowth,SantosD}. ${\mu}$SR experiments under pressure were performed at the GPD instrument (${\mu}$E1 beamline) of the Paul Scherrer Institute (Villigen, Switzerland) \cite{KhasanovPressure}. Time differential and ambient pressure ${\mu}$SR measurements were performed using the General Purpose Surface-Muon Instrument (GPS) \cite{GPS} with a standard low-background veto setup at the ${\pi}$M3 beam line
of the Paul Scherrer Institute in Villigen, Switzerland and using the Los Alamos Meson Physics Facility (LAMPF) spectrometer with a helium gas-flow cryostat at the M20 surface muon beamline (500 MeV) of TRIUMF in Vancouver, Canada.
In a ${\mu}$SR experiment, positive muons ${\mu}$$^{+}$ with nearly 100 ${\%}$ spin polarization are implanted into the sample one at a time. The muons thermalize at interstitial lattice sites, where they act as magnetic microprobes \cite{Kubo,Hayano}. In a magnetically ordered material, the muon spin precesses in the local field $B_{{\rm \mu}}$ at the muon site with the Larmor frequency ${\nu}_{{\rm \mu}}$ = $\gamma_{{\rm \mu}}$/(2${\pi})$$B_{{\rm \mu}}$ (muon gyromagnetic ratio $\gamma_{{\rm \mu}}$/(2${\pi}$) = 135.5 MHz T$^{-1}$). 

Pressures up to 2.0 GPa were generated in a double-walled, piston-cylinder type of cell made of MP35N material, specially designed for ${\mu}$SR experiments \cite{Andreica,KhasanovPressure,GuguchiaNature,GuguchiaNJP}. Daphne oil was used as a pressure transmitting mediumd. The pressure was measured by tracking the superconducting transition of a small indium plate using AC magnetic susceptibility. The sample filling factor of the pressure cell was maximized, resulting in $\sim$40\% of the muons stopping in the sample and the rest in the pressure cell. Therefore, the ${\mu}$SR data in the whole temperature range were analyzed by decomposing the signal into a contribution of the sample and a contribution of the pressure cell according to
\begin{equation}
A(t)=A_SP_S(t)+A_{PC}P_{PC}(t),
\end{equation}
where $A_{S}$ and $A_{PC}$ are the initial asymmetries and $P_{S}$(t) and $P_{PC}$(t) 
are the muon-spin polarizations belonging to the sample and the pressure cell, respectively.
The pressure cell signal was modeled with a damped Kubo-Toyabe function \cite{Maisuradze}.
The response of the sample consists of a magnetic and a nonmagnetic contribution: 
\begin{equation}
\begin{aligned}P_S(t)= \sum_{i=1}^{N}V_{m}\Bigg[{\alpha^{i}e^{-\lambda_{T}^{i}t}\cos(\gamma_{\mu}B_{\mu}^{i}t)}+\beta^{i}e^{-\lambda_{L}^{i}t}\Bigg]\\  
+(1-V_{m})e^{-\lambda_{nm}t}.
\label{eq1}
\end{aligned}
\end{equation}

Here, $N$ = 2 for $x$ = 0 and $N$ = 1 for $x$ ${\geq}$ 0.15. ${\alpha}^{i}$ and ${\beta}^{i}$ = 1 - ${\alpha}^{i}$, are the fractions of the oscillating and non-oscillating ${\mu}$SR signal. $V_{\rm m}$ denotes the volume fraction of the magnetically ordered part of the sample, and $B_{\mu}^{i}$ is the average internal magnetic field at the muon site. ${\lambda_T}^{i}$ and ${\lambda_L}^{i}$ are the depolarization rates representing the transverse and longitudinal relaxing components of the magnetic parts of the sample. The transverse relaxation rate ${\lambda_T}^{i}$ is a measure of the width of the static magnetic field distribution at the muon site and is also affected by dynamical effects (spin fluctuations) \cite{GuguchiaJSNM}.
The longitudinal relaxation rate ${\lambda_L}$ = 1/$T_{1}$ is determined solely by dynamic magnetic fluctuations. ${\lambda_{nm}}$ is the relaxation rate of the nonmagnetic part of the sample. The total initial asymmetry $A_{\rm tot}$ = $A_{\rm S}$(0) + $A_{\rm PC}$(0) ${\simeq}$ 0.285 is a temperature-independent constant. The fraction of muons stopping in the sample ranged between $A_{\rm S}$(0)/$A_{\rm tot}$ ${\simeq}$ 0.40(3) and $A_{\rm S}$(0)/$A_{\rm tot}$ ${\simeq}$ 0.45(3) depending on the pressure, and was assumed to be independent of temperature for a given pressure. The ${\mu}$SR time spectra were analyzed using the open-source software package \textsc{musrfit} \cite{Suter}.

\begin{figure}[t!]
\centering
\includegraphics[width=1.0\linewidth]{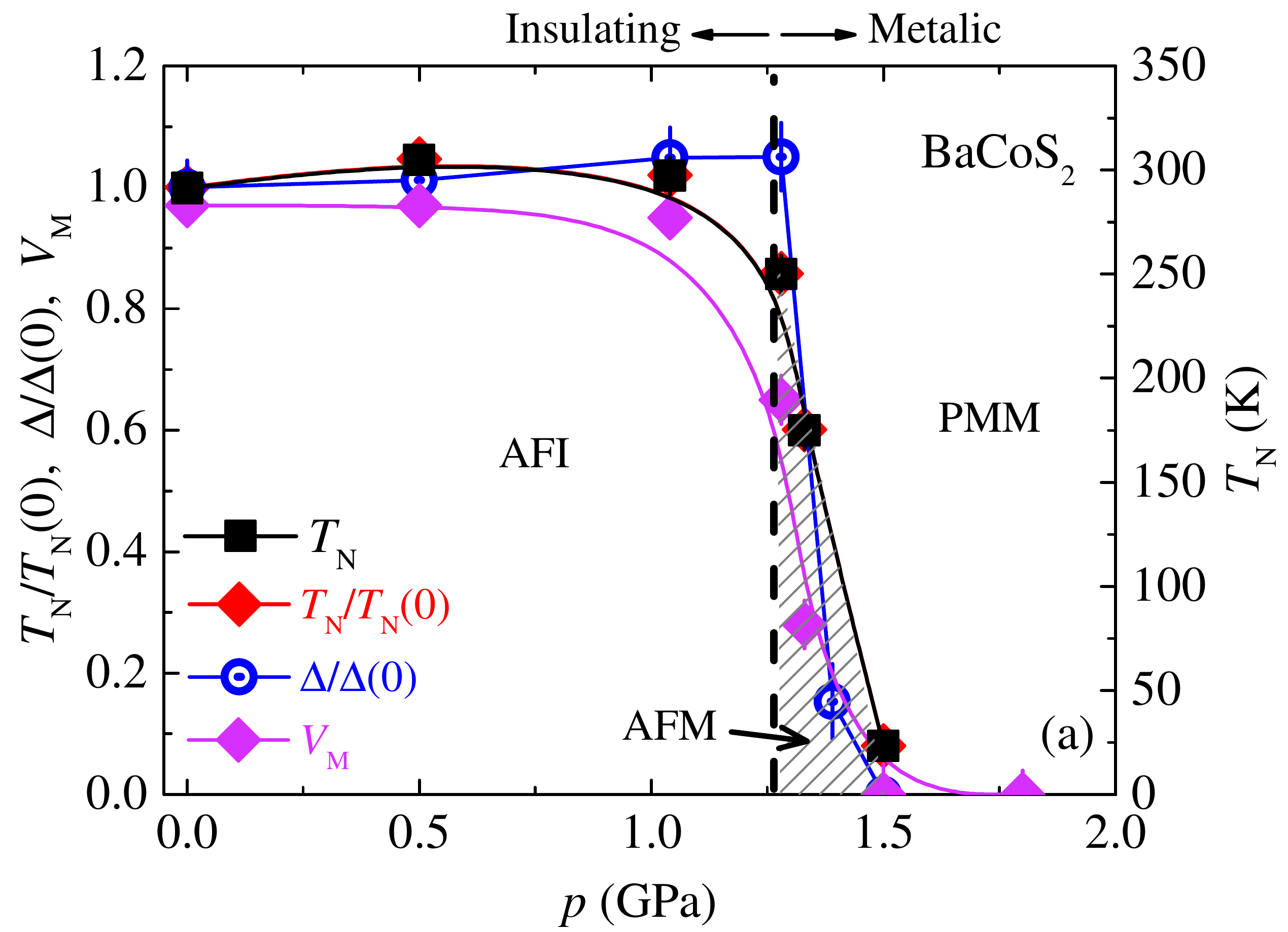}
\includegraphics[width=1.0\linewidth]{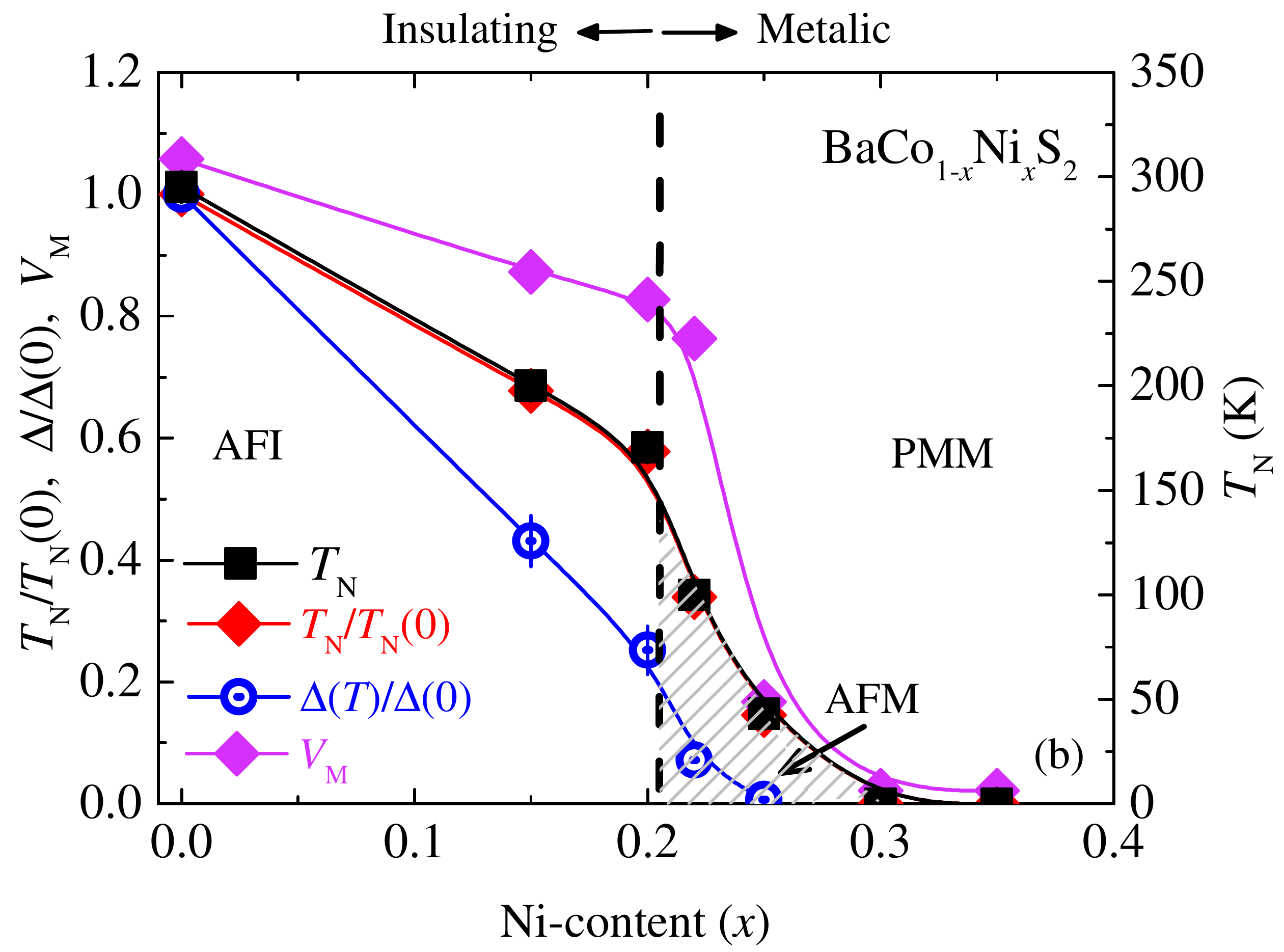}
\vspace{-0.5cm}
\caption{ (Color online) The pressure (a) and doping (b) dependence of the normalized magnetic ordering temperature $T_{\rm N}$/$T_{\rm N} (x = 0, p = 0)$, the normalized zero-temperature static field size ${\Delta}$/${\Delta}(x = 0, p = 0)$, and the zero-temperature magnetically ordered volume fraction $V_{\rm M}$ for BaCoS$_{2}$ under pressure and BaCo$_{1-x}$Ni$_{x}$S$_{2}$, respectively. The broken vertical lines denote the phase boundary between insulating and metallic phases. The shaded area in panel (a) refers to the AF metallic region.}
\label{fig5}
\end{figure}

\begin{figure}[t!]
\centering
\includegraphics[width=1.0\linewidth]{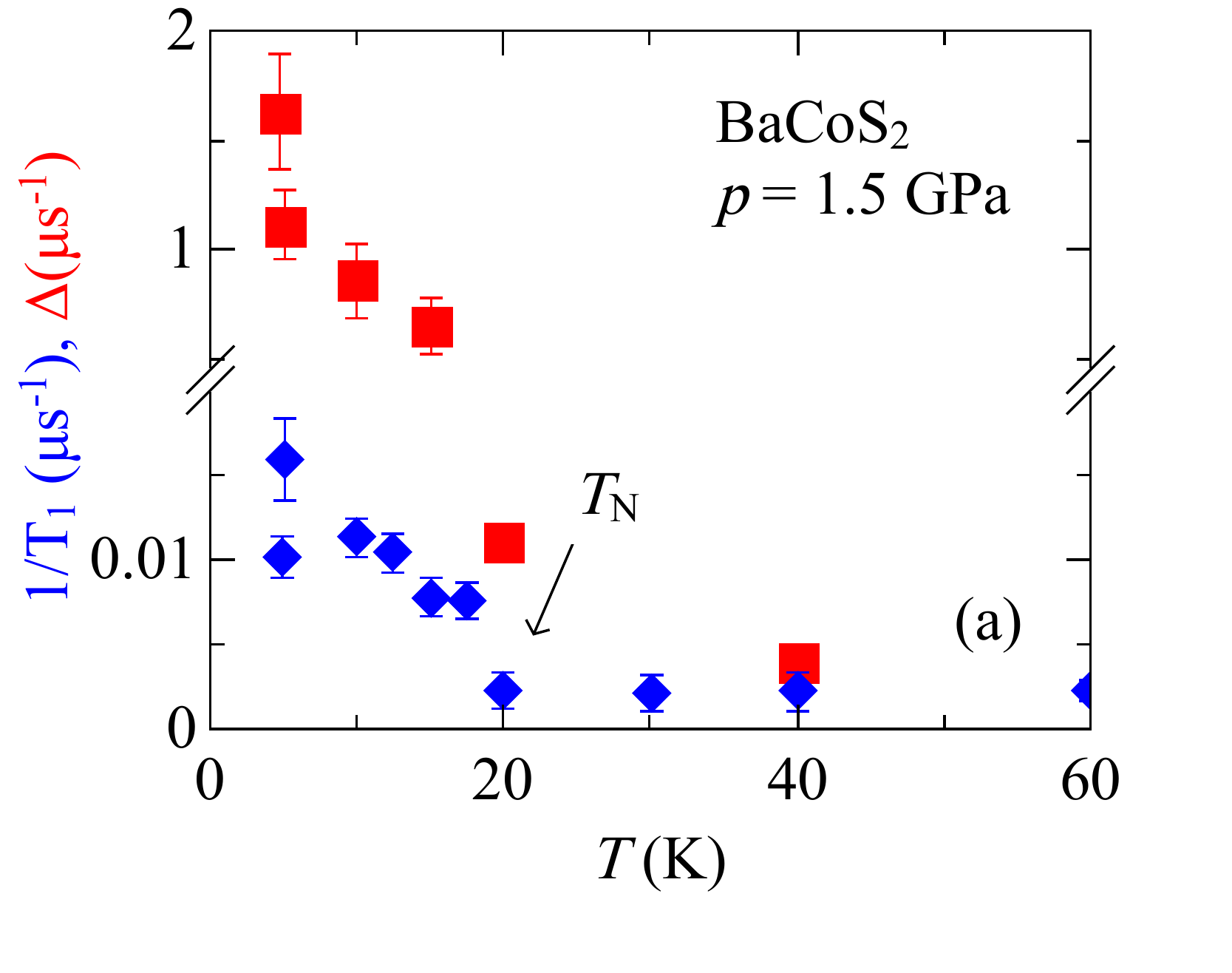}
\includegraphics[width=1.0\linewidth]{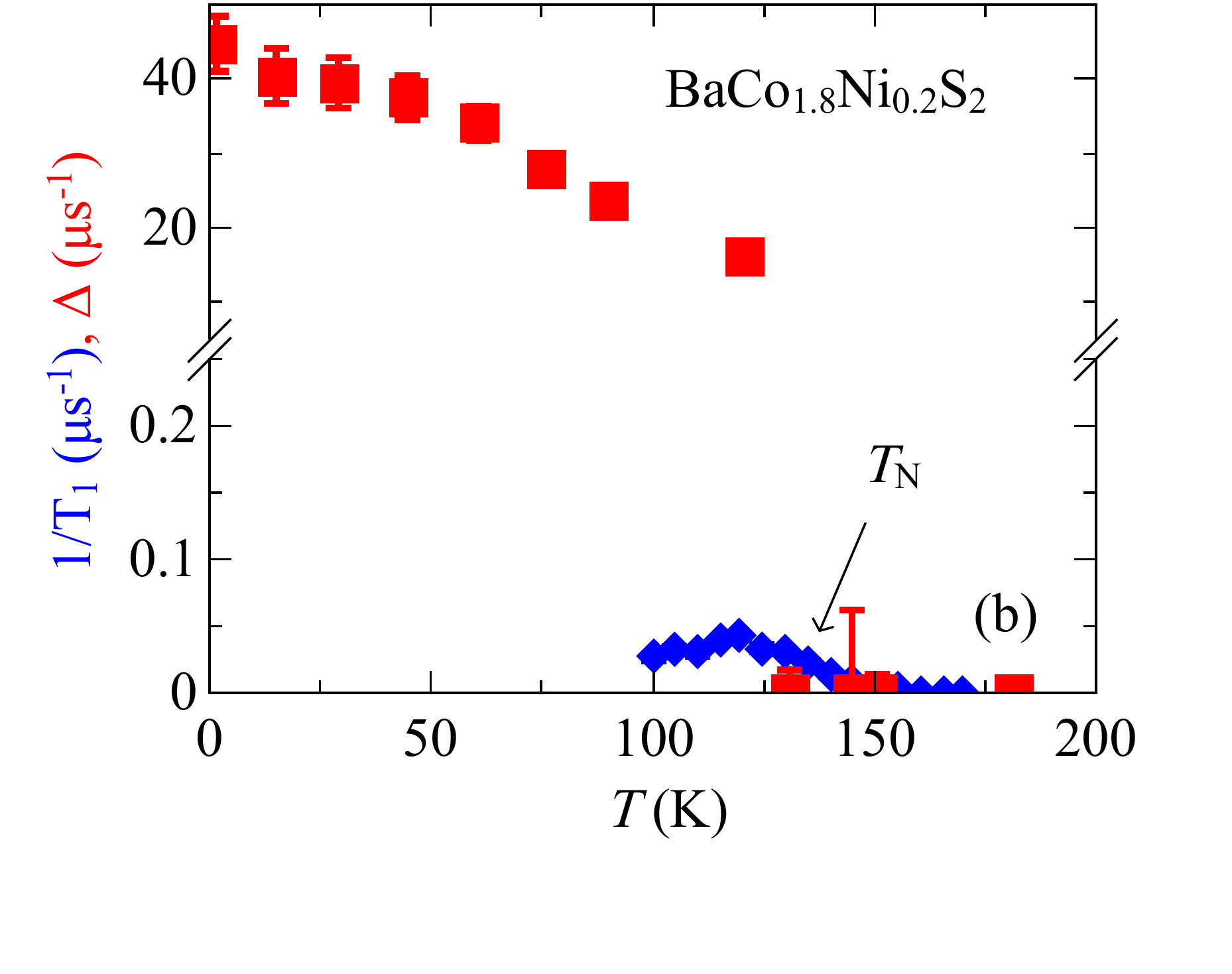}
\vspace{-1.5cm}
\caption{ (Color online) The static field size ${\Delta}$ =[(2${\pi}$${\nu}$)$^{2}$ + ${\lambda_{T}}$$^{2}$)$]^{1/2}$ and the dynamic relaxation rate $1/T_{1}$ for BaCoS$_{2}$ at the applied pressure of $p$ = 1.5 GPa (a) and for the  Ni-doped sample of BaCo$_{1-x}$Ni$_{x}$S$_{2}$ with $x$ = 0.2 (b).}
\label{fig6}
\end{figure}

\section{Results}

We first present the data for pressure-tuned BaCoS$_2$. Fig.~3(a) displays representative zero-field (ZF) ${\mu}$SR time spectra taken at 2 K for various pressures, revealing clear oscillations for pressures up to $p$ = 1.28 GPa. This indicates the existence of a well-defined internal field, as expected in the case of long-range magnetic order. Note that two distinct precession frequencies occur in the ${\mu}$SR spectra, which indicates that two magnetically inequivalent muon stopping sites are present in BaCoS$_2$. Below, we will show and discuss the pressure evolution of the magnetic quantities for one component, since the second component behaves similarly. The ZF oscillation frequency and amplitude are directly proportional to the size of the ordered moment and the magnetically ordered volume fraction, respectively. Slight damping of the oscillations is visible over the time window displayed. In general, damping may be caused by the finite width of the static internal field distribution or by dynamic spin fluctuations. In the present case, the damping is predominantly due to the former effect, as the latter possibility was excluded by applying a longitudinal external field (data not shown). When the system enters the AFM state for $p$ = 1.33-1.5 GPa, we observe relaxation occurring only in a small fraction of the signal with no well-defined oscillations, indicating the existence of disordered static magnetism in a small volume fraction of the sample.

Using the data analysis procedure described above, we obtained quantitative information 
about the oscillation frequency ${\nu}$, the relaxation rates $\lambda_T$ and $\lambda_L$, the magnetic volume fraction $V_{\rm M}$, and the magnitude of the static internal field at the muon site. The latter is calculated as $B_{int} = \Delta / \gamma_{\mu}$, where ${\Delta}$=[(2${\pi}$${\nu}$)$^{2}$ + ${\lambda_{T}}$$^{2}$)$]^{1/2}$. We note that $\Delta$ is proportional to the size of the local ordered moment. Fig.~3(b) displays this quantity for BaCoS$_2$ at various pressures ranging from ambient to 1.5~GPa. Here and elsewhere, we display the fields and the relaxation rates for only one component in the model, since the second component behaves similarly. Up to $p = 1.28$~GPa, $\Delta$ increases continuously from zero as the temperature is decreased below $T_N$ (marked by black arrows), indicative of a continuous phase transition as a function of temperature for these pressures. The value of $\Delta$ at the lowest measured temperature ($\sim$2~K) is virtually unchanged as the applied pressure increases from 0 to 1.28~GPa. However, $\Delta$ is reduced drastically for $p = 1.35$~GPa, at which point the system is in the AFM state. For $p = 1.5$~GPa, $\Delta$ is barely distinguishable from zero.

The magnetically ordered volume fraction is plotted as a function of temperature for several pressures in Fig.~3(c). For $p \le 1.0$~GPa, the magnetically ordered phase at 2~K occupies close to 100\% of the sample volume. This is reduced to 70\% for $p = 1.28$~GPa, close to the AFI - AFM transition at 1.33~GPa, indicating that the ground state of BaCoS$_2$ is characterized by intrinsic phase separation between magnetically ordered and disordered phases at this pressure. The ground-state magnetic volume fraction is further reduced to 30\% for $p = 1.35$~GPa in the AFM phase, and 10\% or less for higher pressures in the AFM phase.

The corresponding results for the Ni-doped system are illustrated in Fig.~4. As seen in Fig.~4(a), long-lived oscillations are observed only in the ZF spectrum for $x=0$, while the spectra for $x \ge 0.15$ display fast relaxation without any oscillations. This ZF relaxation, which can be decoupled in a longitudinal field, points to fairly disordered static magnetism for these high values of $x$. The ZF relaxation rate decreases with increasing $x$, indicating a steady reduction of the internal field at the muon site. This is demonstrated in Fig.~4(b), which displays the temperature dependence of $\Delta$ for $x$ between 0 and 0.3. The low-temperature value of $\Delta$ decreases steadily as $x$ increases, reaching zero around the expected doping level of $\sim$0.25 marking the AFM - PMM transition. This is in contrast to BaCoS$_2$ under pressure, which showed no change in $\Delta$ until an abrupt drop in the AFM state. On the other hand, the behavior of the magnetically ordered volume fraction for Ni substitution seen in Fig.~4(c) is similar to that of the pressure case, showing a steady reduction of the ground-state ordered volume fraction to zero at the AFM - PMM transition. We also confirmed that the end-member compound BaNiS$_{2}$ exhibits no relaxation or oscillations, indicating an absence of static magnetism.

These results can be summarized in Fig.~5, which displays $T_N$, $V_{\rm M}$, and $\Delta$ extrapolated to zero temperature and normalized by its value for pure BaCoS$_2$ at ambient pressure, each as a function of $p$ [Fig.~5(a)] and $x$ [Fig.~5(b)].  With pressure tuning, $\Delta$ (and therefore the ordered moment) displays little change until an abrupt reduction upon entering the AFM state, whereas Ni doping causes a smooth reduction to zero upon entering the PMM state. This behavior, reminiscent of the restoration of a continuous QPT observed in Fe-doped (Mn,Fe)Si tuned by pressure~\cite{Goko}, may be related to effects of disorder \cite{Disorder} introduced by (Co,Ni) substitution. On the other hand, both pressure and Ni doping show a significant reduction of $V_{\mathrm{M}}$ in the vicinity of the QPT, indicating phase separation between antiferromagnetic and paramagnetic regions. Such behavior is consistent with a first-order quantum phase transition, not a continuous transition. For both tuning methods, the AFM state is characterized by a small magnetic volume fraction, weak internal field, and lack of clear oscillations in the ZF spectra, indicative of weak, random, static magnetism in this state.

The first-order versus continuous nature of the QPT can be further clarified by investigating any associated dynamical critical behavior. We performed detailed measurements of BaCoS$_2$ with $p = 1.5$~GPa and Ba(Co$_{1-x}$Ni$_{x}$)S$_{2}$ with $x = 0.2$ in a small longitudinal field (LF) just strong enough to decouple the effect of the static random local field. These values of $p$ and $x$ place the system near the QPT, so any quantum critical dynamics from a continuous QPT would be observable by $\mu$SR. The temperature dependence of the dynamic relaxation rate ${1/T_{1}}$ under these conditions is shown as blue diamonds in Figs.~6(a) and (b) for the pressure- and Ni-tuned samples, respectively, along with the static relaxation parameter ${\Delta}$ in red squares for comparison. For both samples, the magnitude of ${1/T_{1}}$ is less than 1 ${\%}$ of the static relaxation rate  ${\Delta}$($T$=0) and no clear peak is observed at $T_N$. This differs from the case of continuous transitions in (Mn,Fe)Si \cite{Goko} and dilute spin glasses CuMn and AuFe \cite{UemuraGlass}, in which the temperature dependence of ${1/T_{1}}$ exhibits a clear peak at the ordering/freezing temperature with a peak value close to 10 ${\%}$  of the static field parameter \cite{Goko,UemuraGlass}. The absence of dynamical critical behavior in the present data gives further evidence for first-order quantum evolution in BaCoS$_{2}$ for both pressure and doping control.

\section{Discussion}

The $\mu$SR results presented here provide unambiguous evidence for a first-order Mott QPT in BaCoS$_2$ accessed by pressure (bandwidth control) and Ni doping (filling control). This finding is consistent with the recent observation of first-order quantum evolution in the canonical bandwidth-controlled Mott insulators $RE$NiO$_{3}$ and V$_{2}$O$_{3}$, despite the fact that no structural transition accompanies the MIT in BaCoS$_2$. The similarity among disparate Mott systems suggests that first-order quantum phase transitions are ubiquitous in strongly correlated Mott systems, in agreement with previous theoretical predictions \cite{Watanabe,Misawa,Imada,Chitra,Millis,Frandsen2016}.

First-order Mott QPTs such as the one observed in BaCoS$_2$ are reminiscent of the magnetic phase transition observed at the antiferromagnetic/superconducting phase boundary in unconventional superconductors, including high-$T_{\rm c}$ cuprates, FeAs systems, $A_{3}$C$_{60}$, and heavy fermion systems. Similar behavior is also found in $^{4}$He at the boundary between the solid hexagonal close packed (HCP) phase and the superfluid phase \cite{UemuraHe}. In these systems, the superconducting/superfluid phase is accompanied by inelastic excitations associated with short-range correlations having a periodicity characteristic of the competing order (i.e. magnetic or solid HCP). These excitations are referred to as the magnetic resonance mode in unconventional superconductors and rotons in superfluid $^{4}$He \cite{UemuraHe}, as discussed elsewhere~\cite{UemuraHe,UemuraJPCM}. Given the similarities with non-superconducting Mott systems, the universality of inelastic magnetic excitations may extend to BaCoS$_2$, $RE$NiO$_3$, V$_2$O$_3$, and other materials. Therefore, the present results call for further investigations of the role of inelastic soft modes in non-superconducting Mott transition systems.

Another new aspect of the MIT that emerges from the present work is that the AFM region between the AFI and PMM states is characterized by weak, random, static magnetism in a very small volume fraction of the sample. It seems plausible that the magnetic volume is confined to islands embedded in the surrounding PMM phase, suggesting a scenario of metallic conduction through percolation. One open question is the distinction, if any, between the above scenario of separation between AFI and PMM phases and the magnetic phase separation observed in V$_{2}$O$_{3}$ and $RE$NiO$_{3}$ coexisting with insulating bulk conductivity. One possibility is that conductive percolation is achieved in BaCoS$_2$ but not in V$_{2}$O$_{3}$ and $RE$NiO$_{3}$. The length scale and texture of this phase-separated state cannot be probed by ${\mu}$SR, so suitable techniques such as scanning tunneling or magnetic force microscopes and/or spatially-resolved optical probes \cite{Basov} should be used to obtain complementary information in future studies.

\section{Conclusions}

The present ${\mu}$SR results unambiguously demonstrate that the QPT from AFI to PMM in BaCoS$_{2}$ induced by pressure (bandwidth control) and by Ni-doping (filling control) proceeds as a first-order transition without dynamic critical behavior. Upon approaching the QPT from the AFI phase, the magnetically ordered volume fraction decreases steadily until it reaches zero at the PMM phase, resulting in a broad region of electronic phase separation. Sudden destruction of the ordered magnetic moment at the QPT in BaCoS$_{2}$ under pressure and the absence of any dynamical critical behavior in BaCo$_{1-x}$Ni$_{x}$S$_{2}$ further supports the notion of a first-order transition in these materials. Similar behavior is observed in $RE$NiO$_{3}$ and V$_{2}$O$_{3}$, indicating that the basic first-order nature of the Mott transition persists whether or not the MIT is associated with structural phase transition ($RE$NiO$_{3}$ and V$_{2}$O$_{3}$) or the development of an intermediate AFM state (BaCoS$_2$). We expect the present findings to provide further guidance in the quest to  understand the complex process of phase transitions in strongly correlated Mott systems.

\textbf{\section{Acknowledgments}}

The ${\mu}$SR experiments were carried out at the Swiss Muon Source (S${\mu}$S) Paul Scherrer Insitute, Villigen, Switzerland and the TriUniversity Meson Facility (TRIUMF) in Vancouver, Canada. The authors sincerely thank the TRIUMF Center for Material and Molecular Science staff and the PSI Bulk ${\mu}$SR Group for invaluable technical support with ${\mu}$SR experiments. Work at the Department of Physics of Columbia
University is supported by US NSF DMR-1436095 (DMREF), NSF DMR-1610633 and the JAEA Reimei project and a grant from the friends of U Tokyo Inc. Z. Guguchia gratefully acknowledges the financial support by the Swiss National Science Foundation (SNF fellowship P300P2-177832). DSC acknowledges financial support of a PHD grant of the "Emergence program" from Sorbonne Universit\'{e}.


\begin{thebibliography}{22}
\bibitem{Disorder} In the pressure method, the pressure is quite homogeneous, so the whole volume feels the same effect and therefore sees the same environment. In the doping method, the doping is random, so there will be Co and Ni rich local regions, so this would naturally give a range of local dopings and therefore a disorder.
\end{thebibliography}
\end{document}